**REPORT**

# Force- and length-dependent catastrophe activities explain interphase microtubule organization in fission yeast

Dietrich Foethke, Tatyana Makushok, Damian Brunner* and François Nédélec*

Cell Biology and Biophysics, European Molecular Biology Laboratory, Heidelberg, Germany
* Corresponding authors. D Brunner or F Nédélec, Cell Biology and Biophysics, European Molecular Biology Laboratory, Meyerhofstrasse 1, 69117 Heidelberg, Germany. Tel.: +49 6221 387 8597; Fax: +49 6221 387 8512; E-mails: brunner@embl.de or nedelec@embl.de

**The cytoskeleton is essential for the maintenance of cell morphology in eukaryotes. In fission yeast, for example, polarized growth sites are organized by actin, whereas microtubules (MTs) acting upstream control where growth occurs. Growth is limited to the cell poles when MTs undergo catastrophes there and not elsewhere on the cortex. Here, we report that the modulation of MT dynamics by forces as observed *in vitro* can quantitatively explain the localization of MT catastrophes in *Schizosaccharomyces pombe*. However, we found that it is necessary to add length-dependent catastrophe rates to make the model fully consistent with other previously measured traits of MTs. We explain the measured statistical distribution of MT–cortex contact times and re-examine the curling behavior of MTs in unbranched straight *tea1Δ* cells. Importantly, the model demonstrates that MTs together with associated proteins such as depolymerizing kinesins are, in principle, sufficient to mark the cell poles.**
*Molecular Systems Biology* 17 March 2009; doi:10.1038/msb.2008.76
*Subject Categories:* cell and tissue architecture

The fission yeast *Schizosaccharomyces pombe* is a convenient model to study cell morphogenesis (Hayles and Nurse, 2001). Wild-type cells are simple elongated rods growing at the cell poles and dividing in the middle. Yet, previous studies have outlined an interesting interplay between shape, growth and cytoskeletal organization. The first component is the rigid cell wall surrounding yeast cells that maintains cell shape independently of the cytoskeleton. Second, the actin cytoskeleton is essential for cell growth and cell wall remodeling (La Carbona *et al*, 2006). Lastly, although microtubules (MTs) are not required for growth *per se*, they control the location of growth sites by depositing specific marker proteins (Mata and Nurse, 1997; Brunner and Nurse, 2000; Sawin and Snaith, 2004). Abnormal deposition, occurring for example in mutants where MTs are shorter, results in cells that are either bent or branched (Sawin and Nurse, 1998; Snaith and Sawin, 2005). MTs also position the nucleus (Tran *et al*, 2001; Loiodice *et al*, 2005) and thus define the site of cytokinesis (Daga and Chang, 2005; Tolic-Norrelykke *et al*, 2005) and the partitioning of the cell into daughter cells. Hence, by controlling cell growth and division, MTs impact the evolution of shape in the cell lineage. As MTs are constrained within the cell, the converse is also true with MT organization being dependent on cell shape. For the rigid *S. pombe* cells, the two processes occur on very different timescales; with MT lifetimes being in the order of minutes, whereas cells typically double in size after 3 h. Consequently, individual MTs are enclosed in a boundary that is effectively constant during their lifetime. This means that it is valid to first study how MTs depend on cell shape, and to later include cell shape changes. We use here computer simulation for the first step, calculating the dynamic spatial organization of MTs within a fixed cell shape. This approach complements other efforts where cell morphogenesis is modeled with reaction–diffusion equations (Csikasz-Nagy *et al*, 2008) by focusing on the MT cytoskeleton.

Interphase MTs in fission yeast are typically forming 2–6 bundles, which are usually attached to the nucleus at their middle (Tran *et al*, 2001) (Figure 1). Antiparallel MTs overlap at their static minus ends, whereas the plus ends are dynamic and grow from the overlap zone toward the cell poles (Tran *et al*, 2001; Hoog *et al*, 2007). Such bundles transmit forces produced at the cell poles by MT polymerization to the nucleus. To position the nucleus near the middle of the cell (Tran *et al*, 2001; Loiodice *et al*, 2005), MTs should efficiently target the cell poles, and have catastrophes (the switch to depolymerization) that are rare enough to enable MTs to reach the cell cortex but not so rare as to induce bending of the MT around the polar cell wall (Tran *et al*, 2001), a configuration that is not observed in wild-type cells. The most popular explanation for the timing of catastrophes involves multiple molecular activities that are assembled at the cell poles (Mata and Nurse, 1997). Another possibility is that forces caused by MT polymerization feed back on MT dynamics (Dogterom *et al*, 2005). This possibility has so far not been confirmed mainly due to the lack of experimental tools *in vivo*. In this study, we have circumvented this problem using stochastic computer simulations to check whether the effect of force measured *in vitro* can explain the observed MT plus end dynamics in *S. pombe*. This involved constructing three models of MT dynamics: model F (force) in which forces at MT tips regulate MT growth and catastrophe rates, model L (length) in which MT length affects catastrophe rate and model FL which combines both effects.

To simulate the MT cytoskeleton in *S. pombe*, we first idealized its 3D shape as a spherocylinder (see Figure 1B and C). A nucleus and MT bundles were then added and confined within this volume. *In vivo*, MT bundles self-organize (Janson *et al*, 2007), to form overlap zones of 0.84 ± 0.29 µm (Tran *et al*, 2001). For the purpose of this study, it was sufficient to use fixed overlap zones and four bundles, each containing four antiparallel MTs, which is representative of the average situation. The nucleus was represented by a sphere, to which the overlap zones of the bundles were attached (Supplementary Figure S4). The deformations of the nuclear membrane observed *in vivo* (Tran *et al*, 2001; Daga *et al*, 2006) were incorporated into the model by attaching MT bundles to the nucleus using Hookean springs of moderate stiffness. The points to which the bundles were attached were also able to move on the nuclear surface. This allowed elongating bundles to align with the cell axis, as *in vivo* (Supplementary Figure S5). In summary, the simulation comprised bundles of flexible MTs and a connected spherical nucleus that were confined within a frictionless cortex. Although MT minus ends were static, plus ends grew and shrank independently of each other, thus producing polymerization forces, fiber deformation and nuclear movements. The physical equations describing the evolution of this system were solved numerically as explained before (Nedelec and Foethke, 2007). The specific models for this study are detailed in the Supplementary information.

The advantage of using *S. pombe* as a model organism is that numerous dynamical properties of MTs have been measured by light microscopy, and this enables the models to be quantitatively compared to reality. Table I lists the 10 *in vivo* properties of MTs that were used to evaluate computational models. These 10 traits summarize most of the currently established knowledge of the wild-type cells that are relevant for MT organization. Note that it was necessary to adjust the definitions of these traits to what is available from the literature (see Supplementary information). For example, it was important to distinguish 'bundle catastrophes,' that affect the longest MT in a half-bundle from other MT catastrophes. Bundle catastrophes lead to bundle shortening, and are easy to detect experimentally. Other catastrophes are harder to observe because MTs overlap *in vivo*, and were often not reported (Brunner and Nurse, 2000; Drummond and Cross, 2000; Tran *et al*, 2001). Independently of the traits, characteristics of the cell geometry, MT bending elasticity and cytoplasmic viscosity were obtained from the literature or determined experimentally in the course of this study (see Supplementary information). The models discussed here contain two parameters: the polymerization speed $v_0$ and the catastrophe rate $c_0$. They are evaluated by running thousands of simulations with random values of $v_0$ and $c_0$, and then testing which trait was matched in each simulation. This systematic numerical exploration of the parameter space is centered on $v_0 \sim 2.4 \mu m$/min, which is observed under standard laboratory conditions (Tran et al, 2001), and extends to regions where traits start to fail.

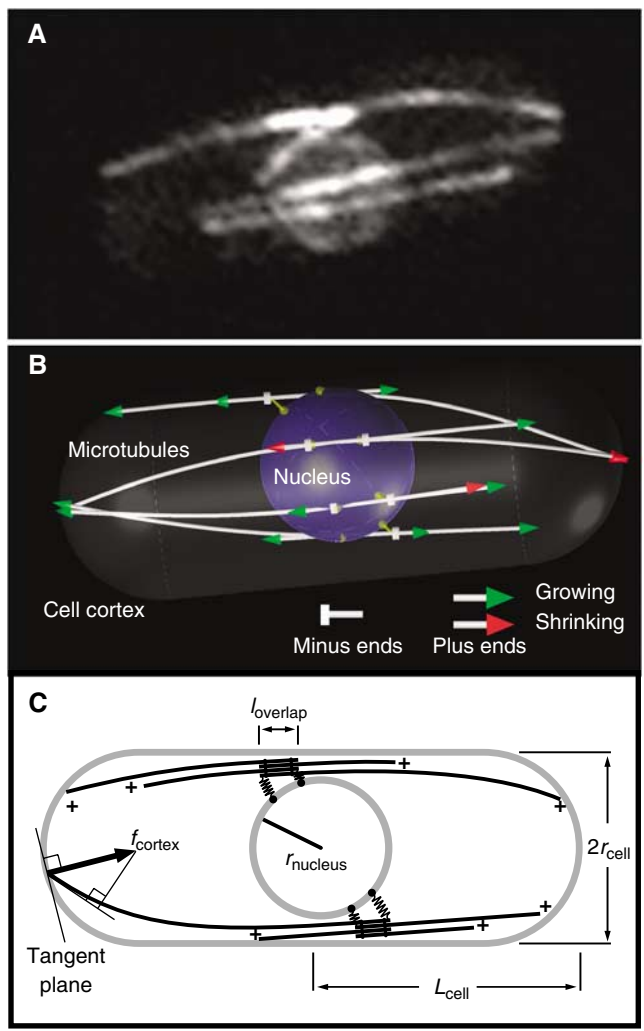

**Figure 1** (**A**) *S. pombe* strain expressing GFP–tubulin and the nuclear pore marker Nup85–GFP. (**B**) The 3D simulation contains a spherical nucleus radius of 1.3 µm and MT bundles attached to it. (**C**) The cell of half-length 5.5 µm is a cylinder closed by half-spheres, of radius 1.6 µm. In each bundle, four MTs overlap near their minus ends, where they are linked to the nucleus. For more information, see Supplementary information.

Table I Measured features of wild-type S. pombe cells (defined in Supplementary information)

| Trait | Description | Range | References |
| --- | --- | --- | --- |
| T1 | Bundle catastrophes at cell poles | 90–100% | Brunner and Nurse (2000) |
| T2 | Bundle catastrophes in contact with the cortex | 90–100% | Brunner and Nurse (2000) |
| T3 | Number of bundles contacting the cell poles | 2–6 | Daga et al (2006) and this study |
| T4 | Mean MT contact time with the cell pole (from contact to catastrophe) | 60–100 s | Brunner and Nurse (2000) and this study |
| T5 | Bundle length divided by cell length | 0.6–1.0 | This study |
| T6 | Probability of seeing a curled MT | 0–1% | Behrens and Nurse (2002) |
| T7 | Variance of nuclear motions | 0–0.25 $\mu m^2$ | Daga and Chang (2005) and Daga et al (2006) |
| T8 | Re-centering speed of initially off-centered nucleus | 0.2–0.9 $\mu m$/min | Daga et al (2006) |
| T9 | Mean contact times of MTs with proximal cell pole, for off-centered nucleus | 20–70 s | Daga et al (2006) |
| T10 | Variance of MT bundles in enucleated cells | 1.4–3.4 $\mu m^2$ | Carazo-Salas and Nurse (2006) |

Model F corresponds to an *in vitro* experiment, which showed that a barrier could inhibit tubulin assembly (Dogterom and Yurke, 1997). These results were described by $v_g = v_0 \exp(-f/f_s)$ (equation A), where $f_s \sim 1.7$ pN specifies the sensitivity to force, and $f$ is the projected force between the barrier and the MT tip (Figure 1C). In addition, reduced assembly *in vitro* was shown to promote catastrophes according to B: $c = 1/(a + bv_g)$ (Janson et al, 2003; Janson and Dogterom, 2004). Model F used equations (A) and (B) with the forces calculated for each MT. The constants $f_s$ and $a$ were measured experimentally with purified tubulin. The constant $b$ was calculated as the solution of (B) with $c = c_0$ and $v = v_0$ such that $c_0$ and $v_0$ correspond to the catastrophe and assembly rates of unconstrained MTs. The resultant value of $b$ differs from that measured *in vitro*, because it represents the combined effects of MT-associated proteins (MAPs) present in the cell. A value for $b$ corresponding to what was reported *in vitro* without MT regulators results in excessively long MTs that curl around the cell pole. We found that simulations with model F consistently matched T1–8 (traits of the wild-type cell) around the reference values of $c_0$ and $v_0$ (Supplementary Figure S1). However, these simulations failed to match T9 and T10 in this region, which lead us to investigate these traits further.

Experimentally, T9 was measured in cells that are made asymmetric by centrifugation, with the nucleus ending up nearer one cell pole. It was observed that during nucleus re-centering, MTs have longer contact time with the proximal than with the distal pole ($\sim 92 \pm 53$ and $\sim 46 \pm 36$ s, respectively; Daga et al, 2006). Model F showed opposite effects because, as shown by our simulation, MTs experience stronger compression on the proximal side than on the distal side. This result agrees with Euler's formula for the critical bucking force $(\pi n/L)^2$, which indicates that the maximum force of an MT at equilibrium is inversely proportional to its length. However, contrary to this formula, the simulation considers the rate of MTs reaching the cell ends, and the motion of the nucleus and thus calculates the effective force rather than an upper limit.

The other unmatched trait T10 represents the centering precision of bundles when they are not attached to the nucleus (Carazo-Salas and Nurse, 2006). To understand why T10 fails, we can consider the simpler situation of an unattached bundle made of two antiparallel MTs. This bundle mechanically behaves as a single elastic beam of small viscous drag: the forces at both ends equalize very fast. Consequently, the plus tips of both MTs experience similar forces no matter where the overlap zone is located: there are little centering cues. The simulation shows that the situation with bundles of four MTs is essentially similar: they are centered on average, but with a large variance of $\sim 3 \mu m^2$. However, when a nucleus is present, additional averaging occurs on the positions of MT minus ends: the nucleus connects multiple independent bundles, and averages their fluctuations over time, because its mobility is much lower than that of individual bundles. Thus, T10 contains different information than T7 and T8. In conclusion, although model F matched most measurements listed in Table I, its failure to match T9 and T10 prompted us to investigate how it could be adjusted.

In the situation considered for T9, MTs are shorter on the proximal side than on the distal side, and therefore the observed asymmetry in contact times may indicate that the *length* of an MT influences its stability, which was not the case in model F. To test this possibility, we simulated a catastrophe rate that depends on MT length $L$ according to $c = hL/(a + bv_g)$ (equation B'). The constant $h$ was set to $0.2/\mu m$, so that $c_0$ would determine the catastrophe rate at $L = 5 \mu m$ (the length at which MTs typically undergo catastrophe in average cells). This makes the resulting model FL directly comparable to model F, except that longer MTs are less stable compared with short ones. With model FL, T9 was matched with, for example, contact times of $102 \pm 68$ s on the proximal side, and $51 \pm 31$ s on the distal side. The length dependence overcompensated for the effects of force in a situation where the nucleus is close to the cell pole. In fact, model FL matched the 10 traits robustly with respect to $(v_0, c_0)$ around their expected *in vivo* values (Figure 2A). Although the average MT–cortex contact time in the symmetric case was already considered in T4, comparing the *shape* of their distribution would provide another test of the model. To this extent, we measured 303 *in vivo* contact events ($66 \pm 36$ s mean and standard deviation). Being approximately 20% shorter, they were comparable to what had previously been reported ($83 \pm 46$ s) (Daga et al, 2006). Interestingly, model FL matched the distribution (see Figure 3A), showing that the contact times are explained by the memory-less (first-order) MT catastrophe transition, because as MTs continue to grow after contact, forces building up increase the instantaneous catastrophe rate. Note that this distribution is also matched by model F (see Supplementary Figure S2), which was anyhow discarded for other reasons.

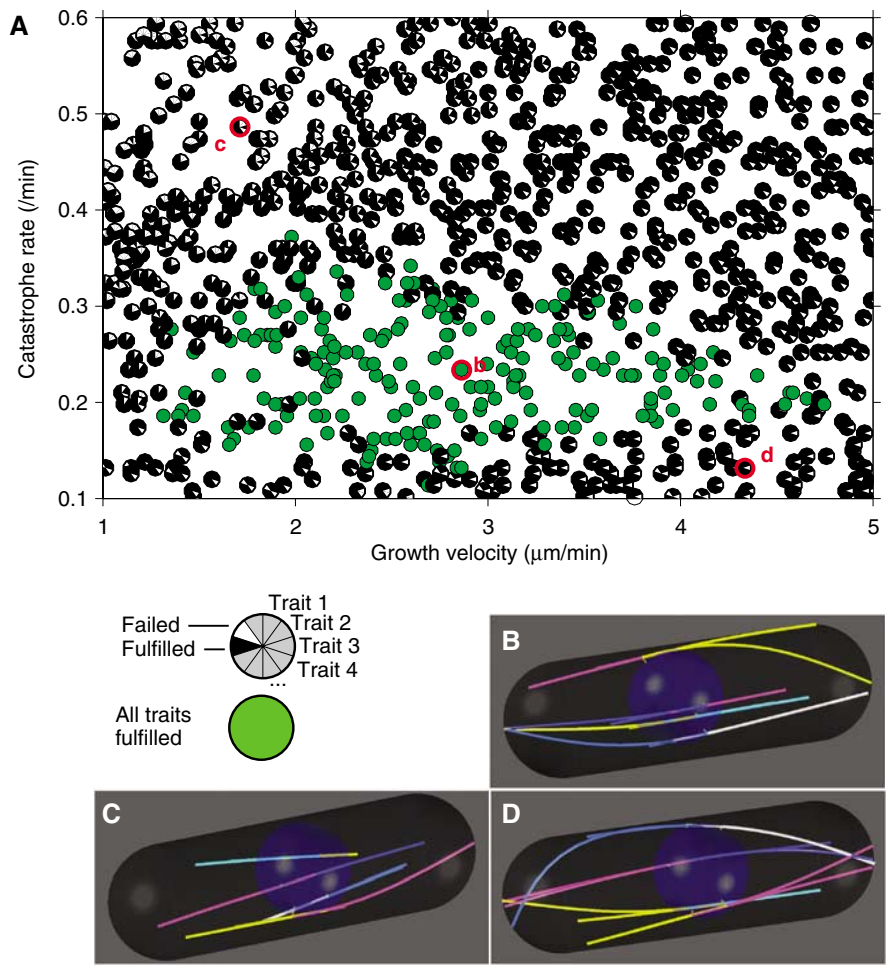

**Figure 2** Trait success–failure diagram. (**A**) To evaluate model FL, we systematically varied the growth speed $v_0$ (*x* axis) and catastrophe rate $c_0$ (*y* axis). Each symbol depicts the outcome of one simulation, summarizing its conformity with the 10 traits of Table I (see legend). (**B–D**) Snapshots from the simulations circled in red in (A).

The simulation also predicts the location of the hidden MT catastrophes (Figure 3B), which are frequent but difficult to observe *in vivo* because these events occur on the side of a longer MT in the same bundle.

Model FL postulated length-dependent catastrophe rates to match all traits. Indeed, such phenomenon was observed previously in *Xenopus* egg extracts (Dogterom *et al*, 1996), and Tischer *et al* have confirmed experimentally that this behavior occurs in *S. pombe* cells, observing that catastrophe rates increase linearly from ∼0.1/min at $L\sim2\,\mu m$ up to ∼0.3/min at $L\sim5\,\mu m$, for MTs that do not contact the cell cortex (Tischer *et al*, in this upload). The measured magnitude of the dependence of catastrophe upon MT length is identical to what we have assumed, in the center of the region in which all traits are fulfilled ($c_0=0.3/min$). It is tempting to speculate about the molecular mechanism that could mediate the destabilization of longer MTs. The *S. cerevisiae* kinesin-8 Kip3p produce a length-dependent destabilizing activity *in vitro*, because after binding at any point along the length of the MT, it moves processively to the tips, where it has a depolymerizing activity (Gupta *et al*, 2006; Varga *et al*, 2006). It is possible that the two homologs of Kip3p in *S. pombe*, Klp5p and Klp6p (West *et al*, 2001) have a similar activity. Consistent with this view, the double deletion of *klp5* and *klp6* produced elongated MTs that curl at cell poles (West *et al*, 2001), and reduces the length dependence of catastrophe (Tischer *et al*). Simulations show that length-dependent catastrophe rates offer several potential advantages to the cell. First, nucleus repositioning is faster by a factor 2 in model FL compared with model F (data not shown), because proximal MTs push for a longer time than distal MTs (T9). Second, bundles that are observed to detach from the nucleus (Tran *et al*, 2001; Carazo-Salas and Nurse, 2006) would keep their overlap regions better centered (their variance was ∼2 μm$^2$ instead of ∼3 μm$^2$ for model F), which may allow them to reattach more rapidly (we did not simulate this hypothesis).

Finally, a model L having the length dependence present in model FL, but without any force dependence also fails to fulfill all traits (Supplementary Figure S1), showing that the length dependence is not sufficient to adjust MTs within the cell. Hence, in model FL, the established response to force together with the length-dependent MT-destabilizing activity of MAPs, leads to the accurate description of the dynamics of MTs in wild-type *S. pombe* cells. Remarkably, in both model F and FL, it was not necessary to assume that the cell poles had any localized activity associated with them that would affect MTs.

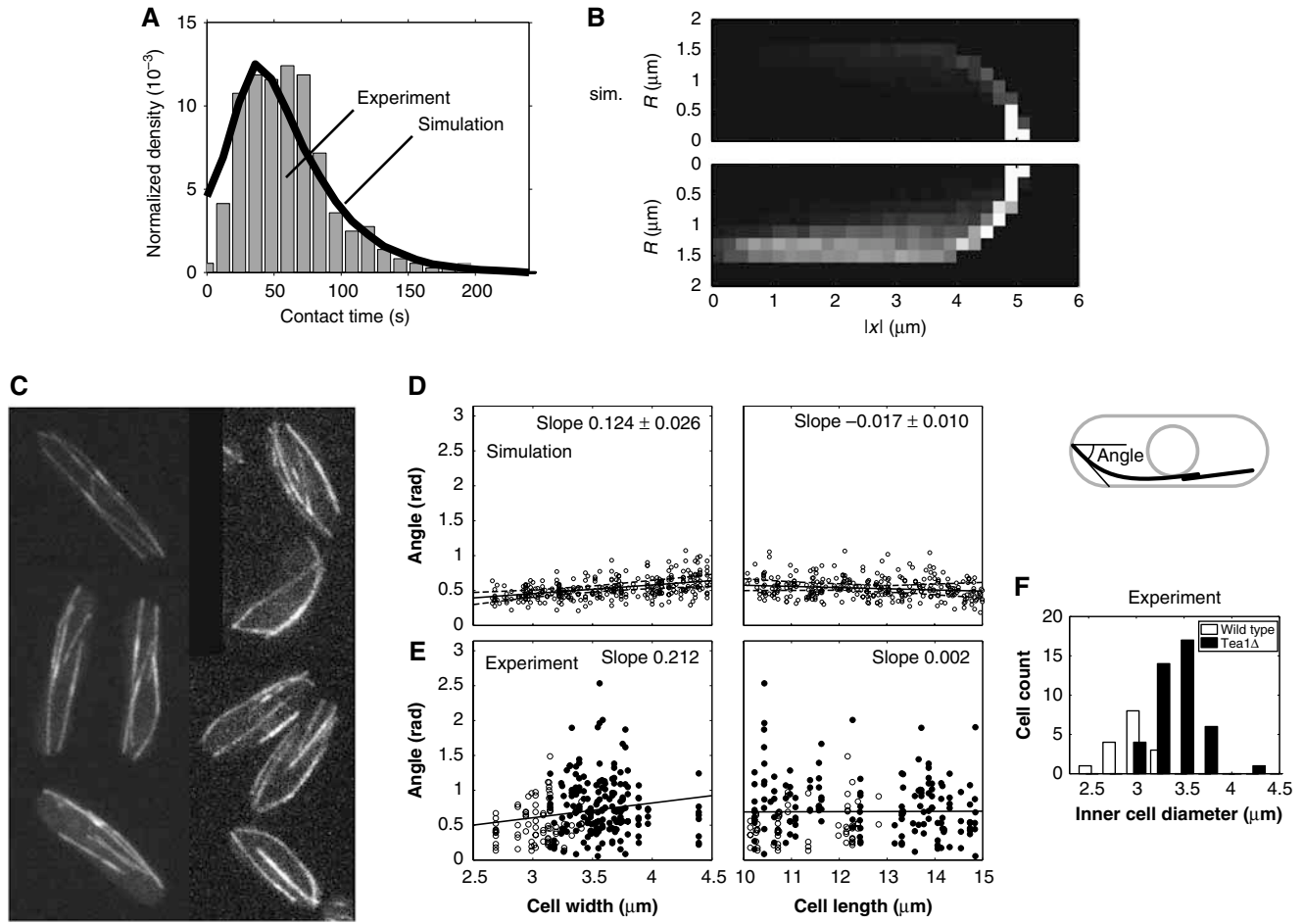

**Figure 3** Simulations compared with experiments. (**A**) MT contact times with the cortex at the cell poles. Bars: histogram of 303 measured events from 343 min of live imaging. Line: distribution predicted by model FL, for the standard parameter values (see Supplementary information). (**B**) Predicted density of catastrophes in the cell, as a function of longitudinal (|x|) and radial ($R=\sqrt{y^2+z^2}$) positions. Top: bundle catastrophes occur mostly at the cell poles as observed. Bottom: hidden catastrophes are more distributed. (**C**) Fluorescence microscopy images of wild-type (left) and *tea1Δ* cells (right) with GFP–α2 tubulin. Compared with wild type, MTs are curling more strongly in *tea1Δ* cells. (**D**) Simulated and (**E**) experimental quantification of the MT curling phenotype: the angle between the cell axis and the MT plus end, at the time of catastrophe, is plotted as a function of cell diameter and cell length. Open symbols on the experimental plot correspond to wild-type cells, and closed symbols to *tea1Δ* cells. The best linear fit to the data is shown, as well as the envelope of many similar fits obtained for other simulation results, which differ only by their random number seed. The slope of the best fit and its standard deviation are indicated. (**F**) Distribution of measured cell widths. Source data is available for this figure at www.nature.com/msb.

This shows that, as anticipated (Dogterom *et al*, 2005), forces in principle are sufficient to account for the location of MT catastrophes at the cell poles.

In future work, it will be necessary to explicitly distinguish MAPs to be able to recapitulate their mutant phenotypes. To do so will require knowing how the dynamic equilibrium distribution of MAPs near MT tips is affected by force. However, having no free parameter, our current model is already predictive in calculating how cell morphology affects a *wild-type* MT cytoskeleton, and this sheds light on some mutant phenotypes. In the current model, the catastrophe rate $c_0$ represents the combined action of all MAPs on MTs. This in particular includes the potential effect of Tea1p, Tip1p or Tea2p, which are located on growing MT plus ends. These proteins are later deposited at the cell ends, but our model does not include any influence of the deposited proteins on MTs in return. In the case of Tea1p, however, such an influence has been suggested (Brunner *et al*, 2000), because genetic deletion affects MTs. The observed defect is that MTs exhibit curling at the cell poles (see Figure 3C). In the simulation, we noticed that cell diameter (but not cell length) affected MT curling (see Figure 3D). This prompted us to measure MT curling *in vivo*, together with the dimensions of cells. There also, we found a clear correlation between cell diameters and the extent of MT curling (see Figure 3E), and no correlation between cell lengths and curling. Thus, the curling phenotype can be explained by the fact that *tea1Δ* cells were, on average, wider than wild-type cells (Figure 3F). Furthermore, thin *tea1Δ* cells exhibited a comparable level of curling to wild-type cells of similar diameter, which confirmed that the MT phenotype is largely the consequence of the increased diameter, and that unlike previously speculated, Tea1p may not directly influence MT catastrophes. The variability around the average behavior is smaller in the simulation than in reality, most likely because irregularities in cell shape together with measurement errors have not been modeled. However, the slopes of the best linear fit were comparable *in vivo* and in the computational model (Figure 3D and E). In summary, it seems that unbranched

*tea1*Δ cells have wild-type MTs in a body that is wider than wild type. It will be important to investigate branched cells to confirm these results. This is beyond the scope of the current study however, because it requires redefining many of the traits, which are only meaningful for cells having two ends.

The simulation available on www.cytosim.org can be extended to further investigate MT organization in *S. pombe*. A successful account of interphase MT dynamics, as initiated here, is a necessary step to understand the determination of cell shape in this simple organism.

## Supplementary information

Supplementary information is available at the *Molecular Systems Biology* website (www.nature.com/msb).

## Acknowledgements

We thank the members of the Nedelec laboratory who contributed to cytosim. We used computers offered by IBM and maintained by the IT support of EMBL. We used microscopes from the EMBL facility. S Huisman, L Murrells and I Aprill helped with the experiments and the writing. We thank M Dogterom, J Gagneur, M Kaksonen, E Karsenti, C Tischer and J Ward for discussions and for critically reading the paper. We gratefully acknowledge HFSP grant RGY84, the Volkswagenstiftung and BioMS for supporting this study.

## Conflict of interest

The authors declare that they have no conflict of interest.

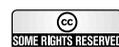